\documentclass[11pt]{article}
\usepackage{comment}
\usepackage[margin=2.5cm]{geometry}
\usepackage{amsmath,amssymb,extarrows,mathtools,graphicx,subfigure,setspace,bbold}
\usepackage{cite}
\usepackage{slashed}
\usepackage[dvipsnames]{xcolor}
\usepackage{hyperref}
\hypersetup{colorlinks=true, linkcolor=blue, citecolor=magenta, linktoc=page}
\numberwithin{equation}{section}
\newcommand{\be}{\begin{equation}}
\newcommand{\bea}{\begin{eqnarray}}
\newcommand{\eea}{\end{eqnarray}}
\newcommand{\ba}{\begin{align}}
\newcommand{\ea}{\end{align}}
\newcommand{\ee}{\end{equation}}
\newcommand{\nn}{\nonumber}

\usepackage{chngcntr}
\counterwithout*{figure}{section}

\begin{document}

\begin{titlepage}
\vspace{10mm}
\begin{flushright}
IPM/P-2018/008 \\

\end{flushright}

\vspace*{20mm}
\begin{center}

{\Large {\bf  Complexity Growth with Lifshitz Scaling \\and Hyperscaling Violation}\\
}

\vspace*{15mm}
\vspace*{1mm}
{Mohsen Alishahiha${}^a$,   Amin Faraji Astaneh$^{b,c}$, M. Reza Mohammadi Mozaffar${}^{a}$ and Ali Mollabashi$^{a}$}

 \vspace*{1cm}

{\it ${}^a$ School of Physics,
Institute for Research in Fundamental Sciences (IPM)\\
P.O. Box 19395-5531, Tehran, Iran\\
 ${}^b$ Physics Department, Faculty of Sciences, Arak University, Arak 38156-8-8349, Iran\\
 ${}^c$  School of Particles and Accelerators,\\
Institute for Research in Fundamental Sciences (IPM)\\
P.O. Box 19395-5531, Tehran, Iran
}

 \vspace*{0.5cm}
{E-mails: {\tt alishah,faraji,m$_{-}$mohammadi,mollabashi@ipm.ir}}%

\vspace*{1cm}
%%\maketitle
\end{center}

\begin{abstract}
Using ``complexity=action'' proposal we study the growth rate of holographic complexity for Lifshitz and hyperscaling violating geometries. We will consider both one and two sided black branes in an Einstein-Maxwell-Dilaton gravitational theory. We find that in either case Lloyd's bound is violated and the rate of growth of complexity saturates to a value which is greater than twice the mass of the corresponding black brane. This value reduces to the mass of the black brane in the isotropic case. We show that in two sided black brane the saturation happens from above while for one sided black brane it happens from below.
\end{abstract}

\end{titlepage}

\newpage
\tableofcontents
\noindent
\hrulefill
\onehalfspacing

\section{Introduction}

In the context of gauge/gravity duality the holographic entanglement entropy \cite{Ryu:2006bv} has provided a geometric (classical) description of a quantum mechanical object. This correspondence might indicate that there could be a deep relation between quantum gravity and quantum information theory. If this is correct, one may expect for instance that 
the nature of space time geometry could be understood from quantum information theory. It would be, then, interesting to understand quantum information theory holographically, in the sense that there could be a holographic dual for some quantum information theory objects.

Actually, the computational complexity could be thought of as an explicit example for this general idea. Indeed based on  earlier works of \cite{{Susskind:2014rva},{Stanford:2014jda}}, it was proposed that for a theory with a gravitational dual, the complexity of a holographic boundary state can be identified with the on-shell action evaluated on a certain subregion of the bulk space time. More precisely, in this proposal which is known as the `complexity=action' (CA), the quantum computational complexity of a holographic state is given by the on-shell action evaluated on a bulk region known as the `Wheeler-De Witt' (WDW) patch \cite{Brown:2015bva, Brown:2015lvg}\footnote{Complexity may also be defined for a subregion
\cite{Alishahiha:2015rta, Ben-Ami:2016qex, Couch:2016exn, Carmi:2016wjl, Bakhshaei:2017qud}.}
\be
{\cal C}(\Sigma)=\frac{I_{\rm WDW}}{\pi \hbar}.
\ee
Here the WDW patch is defined as the domain of dependence of any Cauchy surface in the bulk whose intersection with the asymptotic boundary is the time slice $\Sigma$.

One of the original evidences supporting the proposal is the linear growth of complexity with  respect to time that continues to grow even long after the system reaches thermal equilibrium  \cite{{Susskind:2014rva},{Stanford:2014jda}}. On the other hand for a neutral black hole the growth is bounded by twice of its mass which is sometimes interpreted as the Lloyd's bound on complexity \cite{Lloyd:2000}. Moreover if the dual CFT is perturbed, the corresponding change of complexity matches with holographic complexity in the presence of shockwaves \cite{Stanford:2014jda}.  

To understand complexity and its holographic dual description, it is important to explore different features of it from both holographic and field theoretic points of view. In particular, it is interesting to investigate whether complexity obeys any constraint or any bound such as the Lloyd's bound or not.

Actually despite early observations, it was shown that holographic complexity for the Schwarzschild black hole, which is dual to a thermofield double state, violates the bound \cite{Carmi:2017jqz} (see also \cite{Ghodrati:2017roz,Kim:2017qrq,Moosa:2017yiz}). More precisely although in the late time regime it approaches a constant value that is twice of the mass of the black hole, the constant is approached from above. More recently the complexity growth for a system underlaying a global quench was also studied in \cite{Moosa:2017yvt} where it was shown that the bound is respected during the growth.

The aim of this paper is to further explore Lloyd's bound in a wider family of states supporting both anisotropic and also hyperscaling violating exponents. Such models admit a fixed point where the physics is invariant under an anisotropic scaling   
\be
r\rightarrow \omega r,\;\;\;\;\;\;\;t\rightarrow \omega^z t,\;\;\;\;\;\;\; x\rightarrow \omega x,
\;\;\;\;\;\;\;ds \rightarrow \omega^{\frac{\theta}{d}} ds,
\ee
where $z$ and $\theta$ are anisotropic (Lifshitz) and hyperscaling violating  exponents. Note that with a 
non-zero $\theta$, the distance is not invariant under the scaling which in the context of AdS/CFT 
indicates violations of hyperscaling in the dual field theory. That means in such theories the 
thermal entropy scales as $S_{\rm th}\sim T^{\frac{d-\theta}{z}}$ \cite{{Gouteraux:2011ce},{Huijse:2011ef}}. Holographically the gravity description of these models may be provided by an 
Einstein-Maxwell-Dilaton theory (see for example \cite{{Alishahiha:2012qu},{Salvio:2013jia}}). 

In this paper we will compute the time dependence of holographic complexity for both one and 
two sided black branes in these models using ``complexity = action'' proposal\footnote{
There is also another proposal known as ``complexity=volume'' that we will not consider in this
paper. Actually although in this proposal we have linear complexity growth at late times 
there is no a universal bound for ``CV'' complexity.  Since our main interest is to explore the 
Lloyd's bound in this paper we will only consider ``CA''.}. We note that the late time behavior of 
complexity of such models for two sided black branes has been also studied in \cite{Swingle:2017zcd}.  For a related study in Lifshitz geometry see \cite{An:2018xhv}.

The rest of the paper is organized as follows. In section \ref{sec:2sided} we present the results 
corresponding to two sided black branes where we compute the full time behavior of holographic 
complexity. In section \ref{sec:1sided} we will study one sided black branes. The last section is 
devoted to concluding remarks and some related discussions.

\section{Holographic complexity for  black branes}\label{sec:2sided}

In this section we will compute the on shell action in the WDW patch for black branes with  
Lifshitz and hyperscaling violating exponents. The model that admits such a solution may be given by a 
gravitational theory coupled to a gauge field and a scalar field. The corresponding action is 
\cite{Alishahiha:2012qu}
\bea\label{action}
I=\frac{1}{16\pi G_N}\int d^{d+2}x\sqrt{-g}\left(R-\frac{1}{2}(\partial \phi)^2+V_0 
e^{\xi \phi}-\frac{1}{4}e^{\eta \phi}F^2\right)\,.
\eea
Of course the complete action should have certain Gibbons-Hawking terms defined at space-like 
and time-like boundaries. Moreover to accommodate null boundaries it is also crucial to add the 
corresponding Gibbons-Hawking terms as well as certain joint actions at points of intersection of 
these null boundaries with any other boundary 
\cite{{Parattu:2015gga},{Lehner:2016vdi}}.
Taking all terms into account the action one should consider in the WDW patch is\footnote{
Note that the gauge field is needed to  generate an anisotropy for the metric and therefore 
there is no charge associated to the gauge field. Moreover there is not boundary term for
the gauge field.  See for example \cite{Taylor:2008tg}.}  
\bea
I&=&\frac{1}{16\pi G_N}\int d^{d+2}x\sqrt{-g}\left(R-\frac{1}{2}(\partial \phi)^2+V_0 
e^{\xi \phi}-\frac{1}{4}e^{\eta \phi}F^2\right)+\frac{1}{8\pi G_N}
\int_{\Sigma^{d+1}_t} K_t\; d\Sigma_t\cr &&\cr
&& \pm\frac{1}{8\pi G_N} \int_{\Sigma^{d+1}_s} K_s\; d\Sigma_s\pm \frac{1}{8\pi G_N} 
\int_{\Sigma^{d+1}_n} K_n\; dS 
d\lambda \pm\frac{1}{8\pi G_N} \int_{J^d} a\; dS\,.
\eea
Here the time-like, space-like, and null boundaries and also joint points are denoted by $
\Sigma_t^{d+1}, \Sigma_s^{d+1}, \Sigma_n^{d+1}$ and $J^d$, respectively. The extrinsic 
curvature of the corresponding boundaries are given by $K_t, K_s$ and $K_n$. The function $a$ 
at the intersection of the boundaries is given by the logarithm of the inner product of the 
corresponding normal vectors. $\lambda$ parameterizes the null generator of the null boundary
which in this paper we use Affine parameterization for the null direction. The sign of different terms 
depends on the relative position of the boundaries and the bulk region of interest (see 
\cite{Lehner:2016vdi} for more details).

In what follows we would like to compute the on shell action for black branes with Lifshitz and 
hyperscaling violating exponents given by
\be
ds^2=\frac{L^2}{r_f^{2\frac{\theta}{d}}}\frac{1}{r^{2\frac{d-\theta}{d}}}
\left(-\frac{f(r)}{r^{2(z-1)}}dt^2
+\frac{dr^2}{f(r)}+d\vec{x}^2\right),\;\;\;\;A_t=\frac{L}{r_f^{\frac{\theta}{d}}}\sqrt{\frac{2(z-1)}{d+z-
\theta}}
\frac{1}{r^{d+z-\theta}},\;\;\;\;e^{-\phi}=r^q
\ee
where $L$ is the radius of the geometry, $r_f$ is a dynamical scale where the metric may not be a 
good description for a UV complete theory above it \cite{Dong:2012se}, $q=\sqrt{2(d-\theta)
(z-1-\frac{\theta}{d})}$ and the parameters of the model are given by
\be
\eta=\frac{2\theta(d-1)-2d^2}{q d},\;\;\;\xi=
\frac{2\theta}{ q d}
,\;\;\;V_0=(d+z-\theta-1)(d+z-\theta)\frac{r_f^{2\frac{\theta}{d}}}{L^2}.
\ee
The function $f(r)$ is also given
by
\be
f(r)=1-\left(\frac{r}{r_h}\right)^{d+z-\theta}\,.
\ee
It is worth mentioning that from null energy condition one has \cite{{Dong:2012se},
{Alishahiha:2012qu}} 
\bea\label{NEC}
(z-1)(d+z-\theta)\geq 0,\;\;\;\;\;\;(d-\theta)(d(z-1)-\theta)\geq 0.
\eea
Although from these expressions one could have the possibility of  $\theta>d$, it was
shown that  for this case the solution is unstable  \cite{Dong:2012se}. Therefore in what follows we 
consider $d>\theta$ which  in turn results to $z\geq 1$.

\begin{figure}
\begin{center}
\includegraphics[scale=1.2]{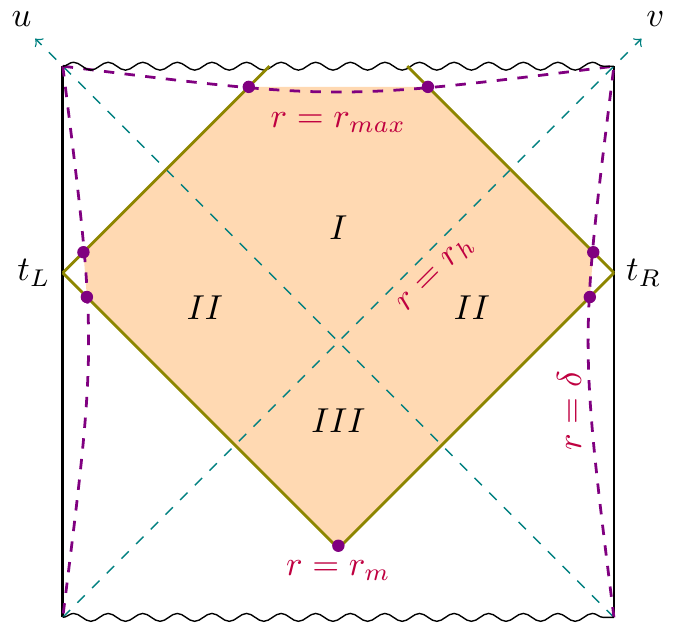}
\end{center}
\caption{WDW patch of a two sided black brane, moving forward in time assuming $t_R=t_L$ .}
\label{fig:A}
\end{figure}

It is useful to define an effective dimension $d_e=d-\theta$, an effective hyperscaling violating 
exponent $\theta_e=\frac{\theta}{d}$ and also an effective scale $L_e=\frac{L}{r_f^{\theta/d}}$. Of 
 course in what follows we set $L_e=1$. In this notation using the trace of Einstein equation 
\be
R-\frac{1}{2}(\partial\phi)^2=-\frac{d+2}{d}V(\phi)+\frac{d-2}{4d}e^{\eta\phi}F^2,
\ee
it is straightforward to see that the action density for the above solution is 
\be\label{bulk}
\sqrt{-g}\left(R-\frac{1}{2}(\partial \phi)^2+V_0 
e^{\xi \phi}-\frac{1}{4}e^{\eta \phi}F^2\right)=-2(1-\theta_e)(d_e+z)
\frac{1}{r^{d_e+z+1}}.
\ee
The null boundaries of the right hand side of the WDW patches (see Fig.\ref{fig:A}) we are 
interested in are
 given by
 \be
 t=t_R+r^*(0)-r^*(r),\;\;\;\;\;\;\;\;\;t=t_R-r^*(0)+r^*(r).
\ee
Actually we should admit that computations of the holographic complexity we will be presenting 
below are  very similar to that of Schwarzschild black hole\cite{Carmi:2017jqz}. In particular due 
to the symmetry of the model the growth rate of complexity is a function of
 $t_L+t_R$ and thus 
for simplicity in what follows we will set $t_L=t_R=\frac{t}{2}$. Moreover as it was shown
\cite{ Brown:2015lvg} (see also  \cite{Carmi:2017jqz} ) there is a critical time $t_c>0$ below which the growth rate of 
 complexity  is zero. Therefore  in what follows we well just present the results for
 $t>t_c$.

To proceed,  using the notation depicted in Fig. \ref{fig:A}, for  a state at 
$t_R=t_L=\frac{t}{2}>\frac{t_c}{2}$  one has\footnote{ The factor of 
 ``2'' is a symmetric factor  due to the symmetry of the WDW patch.} 
 
\bea
I^{\rm bulk}_I&=&-2\times\frac{V_d}{8\pi G_N}(1-\theta_e)(d_e+z)\int_{r_h}^{r_{max}} 
\frac{dr}{r^{d_e+z+1}}\left({\frac{t}{2}+r^*(0)-r^*(r)}\right),\cr &&\cr
I^{\rm bulk}_{II}&=&-2\times\frac{V_d}{4\pi G_N}(1-\theta_e)(d_e+z)\int^{r_h}_{\delta} 
\frac{dr}{r^{d_e+z+1}}\left({r^*(0)-r^*(r)}\right),\cr &&\cr
I^{\rm bulk}_{III}&=&-2\times\frac{V_d}{8\pi G_N}(1-\theta_e)(d_e+z)\int_{r_h}^{r_{m}} 
\frac{dr}{r^{d_e+z+1}}\left({-\frac{t}{2}+r^*(0)-r^*(r)}\right),
\eea
so that
\bea
I^{\rm bulk}&=&-\frac{V_d}{2\pi G_N}(1-\theta_e)(d_e+z)\int^{r_{max}}_{\delta} 
\frac{dr}{r^{d_e+z+1}}\left({r^*(0)-r^*(r)}\right)\cr &&\cr
&&-\frac{V_d}{4\pi G_N}(1-\theta_e)(d_e+z)\int^{r_{max}}_{r_{m}} 
\frac{dr}{r^{d_e+z+1}}\left({\frac{t}{2}-r^*(0)+r^*(r)}\right),
\eea
where $V_d$ is the volume of $d$ dimensional subspace of the space time parametrized by $x_i,
\,i=1,\cdots d$. Also note that $r_m$ can be found from $t=2(r^*(0)-r^*(r_m))$. Note that
at the critical time one has $r_m=r_{\rm max}$, so that $t_c=2(r^*(0)-r^*(r_{\max}))$. 
It is worth mentioning that the time-independent  divergent  term of the bulk action is
\be
I^{\rm bulk}_{\delta}=-\frac{V_d}{2\pi G_N}\,\frac{(1-\theta_e)(d_e+z)}{z d_e \delta^{d_e}}.
\ee

There are also several joint actions from which the only one at $r=r_m$ has non-zero contribution 
to the time dependence of complexity. Taking into account the normal vectors of the corresponding 
null boundaries 
\be
k_1^a={\alpha}\left(\frac{r^{2(z-\theta_e)}}{f}(\partial_t)^a+
{r^{z-2\theta_e+1}} (\partial_r)^a\right),
\;\;\;\;k_2^a={\beta}\left(-\frac{r^{2(z-\theta_e)}}{f}(\partial_t)^a+
{r^{z-2\theta_e+1}} (\partial_r)^a
\right)\,,
\ee
the joint action 
\be 
I^{\rm joint}=\frac{1}{8\pi G_N}\int d^dx \sqrt{\gamma} \;\log \left|\frac{k_1\cdot k_2}{2}\right|,
\ee
reads
\be
I^{\rm joint}=\frac{V_d}{8\pi G_N}\;\frac{1}{r_m^{d_e}}\left((z-\theta_e)
\log {r_m^2}-\log |f(r_m)|\right)+\frac{V_d}{8\pi G_N}\;\frac{\log\alpha\beta}
{r_m^{d_e}},
\ee
where  $\alpha$ and $\beta$ are two constants appearing due to the ambiguity of 
the normalization of normal vectors of null boundaries. On  the other hand from joint points 
at the surface cut off one gets the following time-independent divergent term
\be
I^{\rm joint}_{\delta}=-\frac{V_d}{4\pi G_N}\frac{\log\alpha\beta 
\delta^{2(z-\theta_e)}}{\delta^{d_e}}
\ee
 
There are also several boundaries that could contribute to the time dependence of the action. We 
note, however, that the one at the  time-like cutoff  boundary and the null boundaries do 
not contribute
to the complexity growth rate. The only non-zero contribution comes from the Gibbons-Hawking 
term at the
future  singularity at $r=r_{max}$
\be
I^{\rm surf}=-2\times\frac{1}{8\pi G_N}\int d^dx\, dt \sqrt{h}K_s \Big|_{r=r_{max}}.
\ee
By making use of the fact that
\be
\sqrt{h} K_s=-\sqrt{g^{rr}}\partial_r\sqrt{h}=-\frac{1}{2}\frac{1}{r^{d_e+z-1}}\left(\partial_r f(r)-
\frac{2(d_e+z-\theta_e)}{r}f(r)\right),
\ee
one finds
\bea
I^{\rm surf}&=&\frac{V_d}{8\pi G_N}\frac{1}{r^{d_e+z-1}}\left(\partial_r f(r)-
\frac{2(d_e+z-\theta_e)}{r}f(r)\right)
\left(\frac{t}{2}+r^*(0)-r^*(r)\right)|_{r=r_{max}}\cr &&\cr
&=&\frac{V_d}{8\pi G_N r_h^{d_e+z}}(d_e+z-2\theta_e)
\left(\frac{t}{2}+r^*(0)-r^*(r_{max})\right)\,.
\eea
Since we have used the Affine parametrization for null direction the corresponding boundary
term is zero, while form the time like boundary we get the following divergent term
\be
I^{\rm surf}_\delta=\frac{V_d}{2\pi G_N}\,\frac{d_e+z-\theta_e}{z\delta^{d_e}}\,.
\ee
It is also important to note that due to time like boundary there are certain counter terms 
that needed to make on shell action finite (see {\it e.g.} \cite{Henningson:1998gx}). 
In the present case these terms lead to the following divergent term \cite{TBA}
\be
\hat{I}^{\rm ct}=-\frac{V_d}{2\pi G_N} \,\frac{d_e+z-1}{z\delta^{d_e}}.
\ee
Although this term does not directly contribute to the complexity growth, it is crucial to 
consider it in order to fully understand the structure of counter terms of the model.

Before evaluating the rate of growth  of complexity, we should add proper counter terms to the 
action in order to remove the ambiguity caused by the normalization of null vectors. 
Actually the corresponding counter term that does the work  has the following 
form\cite{Lehner:2016vdi}\footnote{ It is important to note that 
there is an ambiguity for this equation due to a length scale appearing in the logarithm. 
Of course since we have already sent the scale to one, there is no a dimensionful
scale in this expression, though there is still an ambiguity that we fixed it by the factor of
$d_e$ in the   logarithm.  Although  we have fixed the factor by hand there is a way to 
argue how to do that \cite{TBA}.}
 
\be
\frac{1}{8\pi G_N}\int d\lambda d^dx \sqrt{\gamma}\Theta\log\frac{\Theta}{d_e},
\ee
were $\gamma$ is the determinant of the induced metric on the joint point where two null surfaces 
intersect, and 
\be
\Theta=\frac{1}{\sqrt{\gamma}}\frac{\partial\sqrt{\gamma}}{\partial\lambda},
\ee 
with $\lambda$ is an affine parameter for the null surface and in the present  case and for 
the null vector $k_1$ it is given by
\be
\frac{\partial r}{\partial\lambda}=\alpha r^{z-2\theta_e+1}.
\ee
For the null surface associated with the null vector 
$k_1$ one finds $\Theta=-\alpha d_e r^{z-2\theta_e}$ and thus
\bea
I^{\rm ct}_1&=&-\frac{1}{8\pi G_N}\int d\lambda d^dx \sqrt{\gamma}\Theta\log\frac{\Theta}
{d_e}=
\frac{V_d  d_e}{8\pi G_N}\int_\delta^{r_m} \frac{dr}{r^{d_e+1}}\log
{\alpha r^{z-2\theta_e}}\cr 
&&\cr
&=&-\frac{V_d}{8\pi G_N}\left(\frac{\log\alpha r_m^{z-2\theta_e}}{r_m^{d_e}}+\frac{z-2\theta_e}
{d_e r_m^{d_e}}\right)+\frac{V_d}{8\pi G_N}\left(\frac{\log\alpha \delta^{z-2\theta_e}}
{\delta^{d_e}}+\frac{z-2\theta_e}
{d_e \delta^{d_e}}\right),
\eea 
Similarly for the null surface associated with $k_2$ one finds
\bea
I^{\rm ct}_2=-\frac{V_d}{8\pi G_N}\left(\frac{\log\beta r_m^{z-2\theta_e}}
{r_m^{d_e}}+
\frac{z-2\theta_e}
{d_er_m^{d_e}}\right)+\frac{V_d}{8\pi G_N}\left(\frac{\log\beta \delta^{z-2\theta_e}}
{\delta^{d_e}}+\frac{z-2\theta_e}
{d_e \delta^{d_e}}\right).
\eea 
Therefore altogether one gets
\bea
I^{\rm ct}\!&\!=\!&\!-\frac{V_d}{8\pi G_N}\frac{\log\alpha\beta}{r_m^{d_e}}
-\frac{V_d}{4\pi G_N}\frac{\log r_m^{z-2\theta_e}}{r_m^{d_e}}
-\frac{V_d}{4\pi G_N}\frac{z-2\theta_e}
{d_e r_m^{d_e}}\cr &&\cr
&&\!+\frac{V_d}{8\pi G_N}\left(\frac{\log\alpha\beta \delta^{2(z-2\theta_e})}
{\delta^{d_e}}+\frac{2(z-2\theta_e)}
{d_e \delta^{d_e}}\right)\,.
\eea
Note that there are also other counter terms (from null boundaries extended all the way from 
cut off surface $\delta$ to $r_{\rm max}$)  that result to the following divergent term
\be
 {\tilde I}^{\rm ct}= \frac{V_d}{8\pi G_N}\left(\frac{\log\alpha\beta \delta^{2(z-2\theta_e})}
{\delta^{d_e}}+\frac{2(z-2\theta_e)}
{d_e \delta^{d_e}}\right)\,.
\ee
It is then evident that the ambiguous term drops from the on shell action and we would 
also get new time dependent terms contributing to the on shell action. Of course it is not the 
only counter term  that could  make the on shell action finite. Actually one can see 
that there are more counter terms needed to make complexity UV finite\footnote{ We would
like to thank the referee for his/her comment that encouraged us to fully address 
the divergent terms of the model.}. Interestingly 
enough these new counter terms will also lead to new time  dependent terms that  have 
contribution to the complexity growth\cite{TBA}. To explore the situation let us
summarize the terms we have found so far  
\bea
I^{\rm total}\!\!&\!=\!&\!\!I^{\rm bulk}+I^{\rm joint}+I^{\rm surf}+\hat{I}^{\rm ct}+I^{\rm ct}+\tilde{I}^{\rm ct}
\\ \!\! &=&\!\!-\frac{V_d\theta_e}{2\pi G_N}\left(\frac{\log\delta}{\delta^{d_e}}+
\frac{1}{d_e \delta^{d_e}}\right)+\frac{V_d}{2\pi G_N}\frac{z-1}{d_e\delta^{d_e}}+
\frac{V_d}{4\pi G_N}\left(\frac{\log r_m^{\theta_e}}{r_m^{d_e}}-\frac{z-2\theta_e}{d_e r_m^{d_e}}
\right)\cr &&\cr 
\!\!&&\!\!-\frac{V_d}{8\pi G_N}\frac{\log|f(r_m)|}{r_m^{d_e}}+\frac{V_d}{8\pi G_N r_h^{d_e+z}}(d_e+z-2\theta_e)
\left(\frac{t}{2}+r^*(0)-r^*(r_{max})\right)\cr &&\cr
\!\!&&\!\!-\frac{V_d}{4\pi G_N}(1-\theta_e)(d_e+z)\int^{r_{max}}_{r_{m}} 
\frac{dr}{r^{d_e+z+1}}\left({\frac{t}{2}-r^*(0)+r^*(r)}\right)+{\rm finite\, time\, independent\,
term}\,.\nn
\eea
Therefore one needs further counter terms to remove the remaining divergences. Indeed 
one can see that there are certain counter terms that could remove these divergences and 
have  non-trivial contributions  as follows\cite{TBA}\footnote{ Possible counter terms
could be \cite{TBA} $$\frac{1}{8\pi G_N}\int d\lambda d^d\Sigma \sqrt{\gamma}\Theta
\left(\frac{1}{2}\xi\phi+\frac{z-1}{d_e}\right).$$}
\be
 -\frac{V_d}{4\pi G_N}\left(\frac{\log r_m^{\theta_e}}{r_m^{d_e}}+
\frac{\theta_e}{d_e r_m^{d_e}}\right)+\frac{V_d}{4\pi G_N}\frac{z-1}{d_e r_m^{d_e}}.
\ee
Taking these terms into account one arrives at
\bea
I^{\rm total} \!\! &=&\!\!
-\frac{V_d}{4\pi G_N}\frac{1-\theta_e}{d_e r_m^{d_e}}-\frac{V_d}{8\pi G_N}\frac{\log|f(r_m)|}{r_m^{d_e}}+\frac{V_d}{8\pi G_N r_h^{d_e+z}}(d_e+z-2\theta_e)
\left(\frac{t}{2}+r^*(0)-r^*(r_{max})\right)\cr &&\\
\!\!&&\!\!-\frac{V_d}{4\pi G_N}(1-\theta_e)(d_e+z)\int^{r_{max}}_{r_{m}} 
\frac{dr}{r^{d_e+z+1}}\left({\frac{t}{2}-r^*(0)+r^*(r)}\right)+{\rm finite\, time\, independent\,
term}\,.\nn
\eea

Having found all terms contributing to the on shell action, it is then straightforward to compute 
the growth rate of  complexity. Indeed by making use of the fact that $\frac{dr_m}{dt}=
\frac{f(r_m(t))}{2r_m^{z-1}(t)}$ one finds
 \bea\label{R1}
 \frac{d}{dt}{\cal C}=
\frac{1}{\pi }\frac{d}{dt}I_{\rm WDW}=\frac{2E}{\pi}
\left(1+\frac{d_e}{2(d_e+{z-1})}\tilde{f}(r_m(t))\log |f(r_m(t))|\right).
\eea
Here
\be
\tilde{f}(r_m(t))=\left(\frac{r_h^{d_e+z}}
{r_m^{d_e+z}(t)}-1\right),\;\;\;\;\;\;\;\;\;\;
E=\frac{V_d}{16\pi G_N}\;\frac{d_e+z-1}{r_h^{d_e+z}}\,,
\ee
where $E$, at which the complexity approaches at late times,  is a parameter that 
is proportional to the mass of the black brane  (see {\it e.g.} \cite{Dehghani:2015gza})
\be\label{EM}
M=\frac{d_e}{d_e+z-1}E.
\ee
Thus $E$ reduces to $M$ in the isotropic ($z=1$) case. For $\theta=0$ the result
should be compared with that of  \cite{An:2018xhv}\footnote{We note, however, 
that the authors  \cite{An:2018xhv} have not considered the counter terms that remove
the ambiguity and divergences  and therefore their rate of complexity growth has 
unusual behavior for large $z$. See Fig. 4 of  \cite{An:2018xhv}.}.  It is interesting to note that 
despite the fact that the solution depends on $z$ and $\theta$ exponents, qualitatively  the  rate of complexity growth behaves 
the same as that of Schwarzschild black brane\cite{Carmi:2017jqz}. In particular
 it exhibits a logarithmic 
divergence at  times just after the critical time where $r_m\sim r_{\rm max}$
\be
\frac{d}{dt}{\cal C}\sim \frac{2E}{\pi }\left(1
-\frac{d_e(d_e+z)}{2(d_e+z-1)}
\log\frac{r_{\rm max}}{r_h}\right),
\ee
that survives  the $\theta=0$ limit. 
  
\begin{figure}
\begin{center}
\includegraphics[scale=0.82]{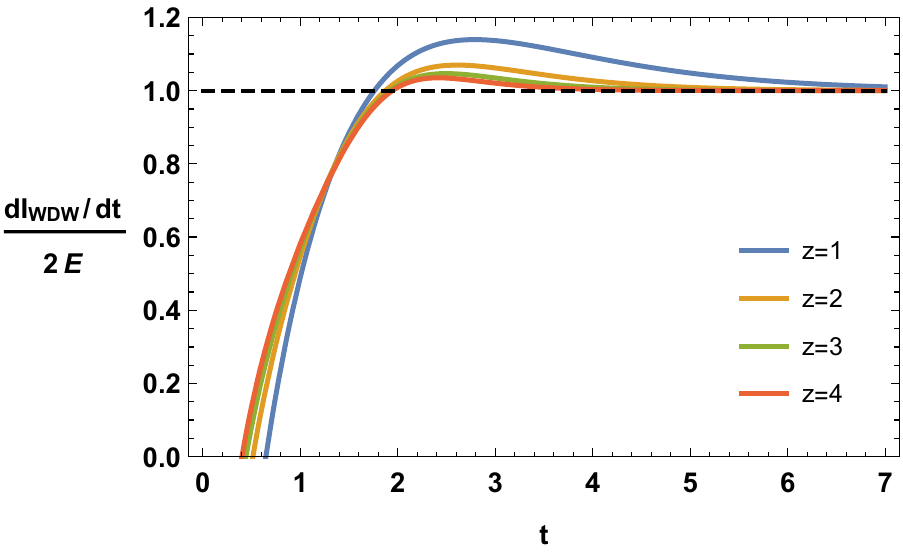}
\hspace{5mm}
\includegraphics[scale=0.82]{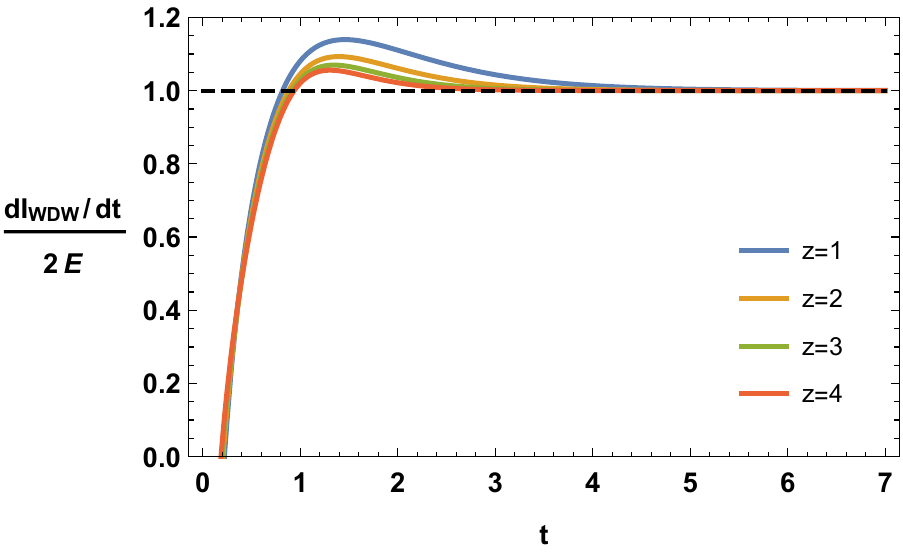}
\end{center}
\caption{Rate of the complexity growth in a WDW patch for two sided black brane. Left (right) panel 
shows different values of dynamical exponent for $d_e=1 (d_e=2)$. For each curve on both 
panels the Lloyd's bound is violated at some time before the curve reaches unity on the vertical 
axes.}
\label{fig:2s}
\end{figure}

On the other hand this result shows that Lloyd's bound (defined in terms of  the mass of black 
brane) is always violated for non-trivial anisotropic and hyperscaling violating exponents. This is 
simply because that  the value of $E$,  at which the rate of complexity growth saturates to, is 
always greater than  (or equal to)  the mass of  the black brane $M$\footnote{
Note that from equation \eqref{NEC} and with the assumption of $d>\theta$ one has
 $z\geq1$.}, which naturally appears on the right hand side of Lloyd's inequality. Of course one 
 may wonder that due to non-trivial scaling of the time coordinate, the Lloyd's bound gets modified 
 from $2M$ to a ``would be" bound $2E$.  We note, however, that even this ``would be" bound is 
 also violated in the present case simply because  the rate of complexity growth approaches the 
 bound given by $2E$ from above (note that  $r_m(t)\geq r_h$).
More precisely at late times where $r_m$ approaches the radius of horizon $r_h$  one gets
\bea
\frac{d}{dt}{\cal C}\sim \frac{2E}{\pi }\left(1+\frac{d_e(d_e+z)}{2(d_e+z-1)}
\left(1-\frac{r_m(t)}{r_h}\right)
\log(d_e+z){|1-\frac{r_m(t)}{r_h}|}\right).
\eea  
The behavior of the rate of complexity growth is depicted in Fig.\ref{fig:2s} for different values of $z$ and $\theta$.  
%\begin{figure}
%\begin{center}
%\includegraphics[scale=0.82]{hs2s1.pdf}
%\hspace{5mm}
%\includegraphics[scale=0.82]{hs2s2.pdf}
%\end{center}
%\caption{Rate of the complexity growth in a WDW patch for two sided black brane. Left panel 
%shows different values of dynamical exponent for $d=1$ and $\theta=0$. The right panel shows 
%different values of hyperscaling violating exponents for $d=3$ and $z=1$. For each curve on both 
%panels the Lloyd's bound is violated at some time before the curve reaches unity on the vertical 
%axes.}
%\label{fig:2s}
%\end{figure}

It is worth noting that whatever the bound is ($2M$ or $2E$)  it is evident from the expression
 \eqref{R1}  that the rate of growth of complexity reaches the  bound  at a finite time which is of 
 order of $r_h$. Of course after this time the complexity still evolves with time till it reaches the bound once again at late times. 

By making use of the equation \eqref{R1} one can find the point  at which the bound 
is being saturated. This can be done by setting the time dependent part of the equation \eqref{R1} to zero
\bea\label{sat}
\tilde{f}(r_m)\log|f(r_m)|=0,
\eea
that solves for $r_m=2^{\frac{1}{d_e+z}}r_h$. Note that it also approaches zero for
$r_m\rightarrow r_h$ that occurs at late times. It is also worth noting that  from the explicit 
form of the blacking function $f(r)$ one gets
\be
t=\frac{2r_m^z}{z}{}_2F_1\left(1,\frac{z}{d_e+z},1+\frac{z}{d_e+z},(\frac{r_m}{r_h})^{d_e+z}\right),
\ee
that may be used to read the corresponding time of the special  points. In particular 
for $r_m\rightarrow r_h$ one has $t\rightarrow \infty$ (late times) and for  $r_m=0$ one gets
$t=0$.  Moreover the critical time $t_c$ where $r_m\rightarrow \infty$ is also given by
\be
t_c\sim \frac{1}{4\sin\frac{z\pi}{d_e+z}}\,\frac{1}{T}\,,
\ee
where $T$ is the Hawking temperature of the black hole. 
%%%%%%%%%%%%%%%%%%%%%%%%%%%%%%%%%%%%%%%%%%%%%%
%%%%%%%%%%%%%%%%%%%%%%%%%%%%%%%%%%%%%%%%%%%%%%
\section{Holographic complexity for Vaidya metric}\label{sec:1sided}

In this section we will consider holographic complexity for Vaidya geometries with Lifshitz and hyperscaling violating exponents. The model could provide a gravity description for a global quench in a field theory with anisotropic scaling and hyperscaling violation. Adding an infalling 
null shell matter to the action \eqref{action}, the resultant model admits the following Vaidya metric  \cite{Alishahiha:2014cwa}
\be
ds^2=\frac{1}{r^{2(1-\theta_e)}}\left(-\frac{f(r,v)}{r^{2(z-1)}}dv^2
-\frac{2}{r^{z-1}}dr dv+d\vec{x}^2\right),\;\;\;\;A_v=\sqrt{\frac{2(z-1)}{d_e+z}}
\frac{1}{r^{d_e+z}},\;\;\;\;e^{-\phi}=r^\beta,
\ee
where 
\bea
f(r,v)=\Bigg\{ \begin{array}{rcl}
&1&\,\,\,v<0,\\
&1-\left(\frac{r}{r_h}\right)^{d_e+z}\equiv f(r)&\,\,\,v>0,
\end{array}\,\,.
\eea
All parameters of the solution are exactly the same as the black brane solution presented in the 
previous section.

\begin{figure}
\begin{center}
\includegraphics[scale=0.8]{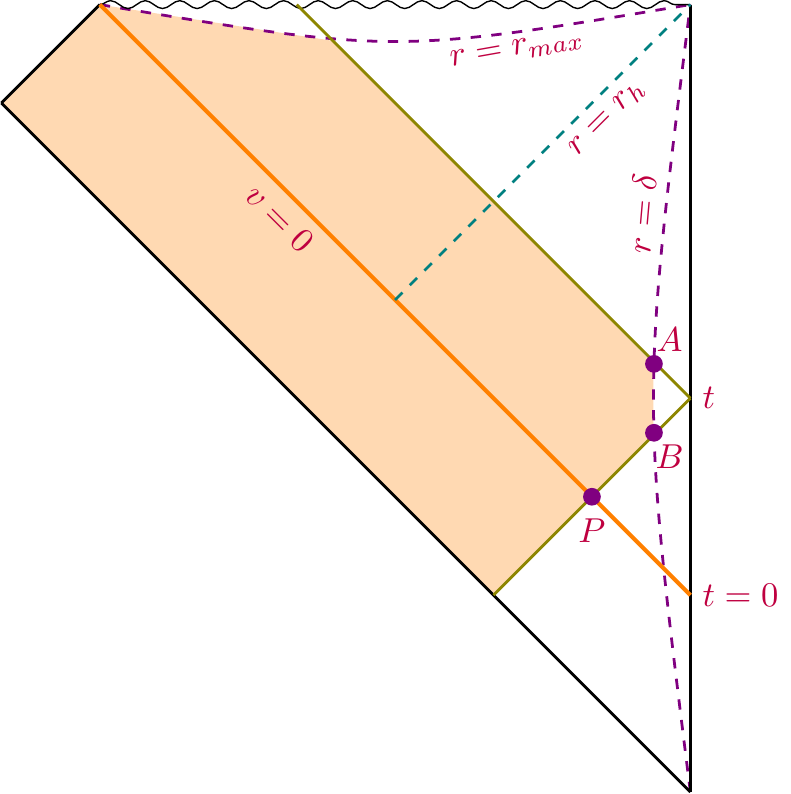}
\end{center}
\caption{WDW patch of a one sided black brane, i.e., Vaidya geometry. The orange line is the in-falling collapsing null shell located at $v=0$.}
\label{fig:C}
\end{figure}

The aim of this section is to compute complexity for the boundary state at time equal to $t$. To do 
so, one needs to compute the on shell action in the corresponding WDW patch depicted in 
Fig. \ref{fig:C}. Following \cite{Moosa:2017yvt} we decompose the patch into two parts: $v>0$ and 
$v<0$ parts (see Fig. \ref{fig:C}).

\subsection*{$v>0$ part}
For $v>0$ we have five boundaries, one at the future singularity and three null boundaries 
given by $v=0$, $v=t$ and 
\be 
t-v=2\int_0^{r_p(v,t)}dr\frac{r^{z-1}}{f(r)},
\ee
and a time like boundary at $r=\delta$.
Using the explicit expression for $f(v,r)$ one finds
\bea\label{tv}
t-v=2\frac{r_p^z}{z} \,{}_2F_1\left(1,\frac{z}{d_e+z},1+\frac{z}{d_e+z},
\left(\frac{r_p}{r_h}\right)^{d_e+z}\right).
\eea
Using the notation of \cite{Moosa:2017yvt}  the  coordinates of the points $A$, $B$ and $P$ are determined 
as follows\footnote{Note that in order to fix $v_B$ we expand \eqref{tv} at 
$r_p\sim \delta\rightarrow 0$
$$
t-v_B=2\frac{\delta^z}{z} \,{}_2F_1\left(1,\frac{z}{d_e+z},1+\frac{z}{d_e+z},
\left(\frac{\delta}{r_h}\right)^{d_e+z}\right)\sim 2\frac{\delta^z}{z}.$$}
\bea
A:&&\;\;\;\;v_A=t,\;\;\;\;\;\;\;\;\;\;\;\;\;\;\;r_A=\delta,\nonumber\\
B:&&\;\;\;\;v_B=t-2\frac{\delta^z}{z},\;\;\;\;\;r_B=\delta,\nonumber\\
P:&&\;\;\;\;v_P=0,\;\;\;\;\;\;\;\;\;\;\;\;\;\;\;r_P=r_p(0,t).
\eea
It is then easy to compute the bulk action in this part. Indeed by making use of 
equation \eqref{bulk} one has
\bea
I^{\rm bulk}_{v>0}\!\!&\!=\!\!&\!-\frac{1}{8\pi G_N}(1-\theta_e)(d_e+z)
\int d^{d}x\;\bigg( \int_0^{v_B} 
dv\int_{r(v,t)}^{\infty} \frac{dr}{r^{d_e+z+1}}+ \int_{v_B}^t dv\int_{\delta}^{\infty} 
\frac{dr}{r^{d_e+z+1}}\bigg)\nonumber\\
\!&\!=\!&\!-\frac{V_d}{8\pi G_N}(1-\theta_e)\left(\frac{2}{z\delta^{d_e}}
+\int_0^{t-\frac{2}{z}\delta^z}\frac{dv}
{r^{d_e+z}(v,t)}\right)\,.
\eea
The $v>0$ part of the WDW has five boundaries three of which are null and their contributions to 
the on shell action vanish using affine parametrization for the null directions. There is also a 
time-like boundary at the cutoff surface and one needs to consider the corresponding 
Gibbons-Hawking term for this boundary. We note, however, that since our ultimate goal is to 
compute the time dependence of complexity, this term does not contribute to the complexity 
growth. On the other hand the space-like boundary at the future singularity does indeed contribute 
to the complexity growth and therefore we will compute the corresponding Gibbons-Hawking term 
given by
\bea
I^{\rm max}_{v>0}=-\frac{1}{8\pi G_N} \int_{\Sigma^{d+1}_s} K_s\; d\Sigma_s=-\frac{1}{8\pi G_N} 
\int 
d^dx\int_0^t dv  \sqrt{\gamma}K_s\, .
\eea
From  the explicit form of the normal vector to the space-like boundary at the future singularity
\bea
n^a=-\frac{r_h^{\frac{d_e+z}{2}}}{r_{\rm max}^{\frac{d_e+z}{2}+{\theta_e}}}(\partial_v)^a-
\frac{r_{\rm max}^{\frac{d_e+z}{2}+1-\theta_e}}{r_h^{\frac{d_e+z}{2}}}(\partial_r)^a,
\eea 
the above Gibbons-Hawking term reads
\bea
I^{\rm max}_{v>0}=\frac{V_d}{16\pi G_N}\frac{d_e+z-2\theta_e}{r_h^{d_e+z}}t\,.
\eea
Finally we need to compute the contributions of the joint points where a null boundary intersects 
with another boundary (that could also be a null one). The corresponding term is given by the 
logarithm of the inner product of two intersecting boundaries. To compute such terms one needs to 
find the normal vectors to the null and time-like boundaries that are given by
\bea
&&{\rm At}\;v=0,t\;\;\;\;\;\;k_1^a=-\alpha\, r^{z+1-{2\theta_e}}(\partial_r)^a,\cr &&\cr
&&{\rm At}\; BP\;\;\;\;\;\;\;\;\;\;\;\;k_2^a=\beta\left(\frac{2}{f(r)}r^{2(z-{\theta_e})}(\partial_v)^a-
r^{z+1-{2\theta_e}}(\partial_r)^a\right),\cr &&\cr
&&{\rm At}\;r=\delta,\;\;\;\;\;\;\;\;k_3^a=r^{z-{\theta_e}}(\partial_v)^a-r^{1-{\theta_e}}
(\partial_r)^a\,.
\eea
It is straightforward to see that the joint terms at $r=\delta$ (at points $A$ and $B$) do not 
contribute to the time dependence of the on shell 
action\footnote{The corresponding contribution is 
$$S= \frac{V_d}{8\pi G_N}\,\left(\frac{\log\delta}{\delta^{d_e}}-\frac{1}{\delta^{d_e}}\right)$$.}
while the one at point $P$ does. Therefore in what follows we only consider the joint 
term at pint $P$ where two null boundaries intersect. Indeed by making use of these 
normal vectors one can find the contribution of the joint point at $P$ as follows
\bea
I^{\rm joint }_{v>0}&=&\frac{1}{8\pi G_N}\int_p \sqrt{\gamma}\; d^dx  \log |\frac{k_1\cdot k_2}{2}|
\cr &&\cr
&=&\frac{V_d}{8\pi G_N} \frac{1}{r_p^{d_e}(t)}
\left((z-\theta_e)\log r^2_p(t)-\log f(r_p(t))\right)+\frac{V_d}{8\pi G_N} 
\frac{\log\alpha\beta}{r_p^{d_e}(t)}.
\eea
Putting all results together one gets
\bea
I_{v>0}\!&\!=\!&\!\frac{V_d}{8\pi G_N}(1-\theta_e)\left(
\frac{\log\delta}{\delta^{d_e}}-
\frac{2}{z\delta^{d_e}}
-\int_0^{t-\frac{2}{z}\delta^z}\frac{dv}
{r^{d_e+z}(v,t)}\right)
+\frac{V_d}{16\pi G_N}\frac{d_e+z-2\theta_e}{r_h^{d_e+z}}t\cr &&\cr
&&+\frac{V_d}{8\pi G_N} \frac{1}{r_p^{d_e}(t)}
\left((z-\theta_e)\log r^2_p(t)-\log f(r_p(t))\right)+\frac{V_d}{8\pi G_N} 
\frac{\log\alpha\beta}{r_p^{d_e}(t)}.
\eea

\subsection*{$v<0$ part}
In this part the function $f(v,t)=1$ and we have four boundaries two of which are given by the past and future horizon and two null boundaries located at $v=0$ and 
\be
r^z=r^z_p(t)-\frac{z}{2}v.
\ee 
It is then straightforward to compute the bulk action in this region
\bea
I^{\rm bulk}_{v<0}&=&-\frac{1}{8\pi G_N}(1-\theta_e)(d_e+z)\int d^{d}x\;
\int_{-\infty}^0 dv\; \int_{(r_p^z(t)-\frac{z}{2}v)^{1/z}}^\infty \frac{dr}{r^{d_e+z+1}}\cr &&\cr
&=&-\frac{V_d}{4\pi G_N}\frac{1-\theta_e}{d_e}\;\frac{1}{r^{d_e}_p(t)}.
\eea
The contribution of joint point at $P$ is 
\be
I^{\rm joint }_{v<0}=-\frac{1}{8\pi G_N}\int_p \sqrt{\gamma}\; d^dx  \log |\frac{k_1\cdot k_2}{2}|,
\ee
where 
\be
k_1^a=-\alpha\,r^{z+1-{2\theta_e}}\partial_r,\;\;\;\;\;\;\;\;
k_2^a=\beta\left({2}r^{2(z-\theta_e)}\partial_v-
r^{z+1-{2\theta_e}}\partial_r\right)\, .
\ee
Therefore one finds
\be
I^{\rm joint }_{v<0}=-\frac{V_d}{8\pi G_N} 
\frac{z-\theta_e}{r_p^{d_e}(t)}\log r^2_p(t)-\frac{V_d}{8\pi G_N} 
\frac{\log \alpha\beta}{r_p^{d_e}(t)}.
\ee
Since the null boundaries do not contribute to the on shell action, taking both bulk and joint 
contributions into account one 
arrives at
\be
I_{v<0}=-\frac{V_d}{8\pi G_N}\;\frac{1}{r^{d_e}_p(t)}
\left(\frac{2(1-\theta_e)}{d_e}+(z-\theta_e) \log r^2_p(t)\right)-\frac{V_d}{8\pi G_N} 
\frac{\log \alpha\beta}{r_p^{d_e}(t)}.
\ee 

\subsection*{On shell action and complexity growth}
Now we have all ingredients to write the on shell action on the WDW patch for our Vaidya 
geometry. Indeed one has
\bea
I=I_{v>0}+I_{v<0}\!&\!=\!&\!-\frac{V_d}{8\pi G_N}(1-\theta_e)
\int_0^{t-\frac{2}{z}\delta^z}\frac{dv}
{r^{d_e+z}(v,t)}+\frac{V_d}{16\pi G_N}\frac{d_e+z-2\theta_e}{r_h^{d_e+z}}t
\cr &&\cr && -\frac{V_d}{4\pi G_N} \frac{1}{r_p^{d_e}(t)}
\left(\frac{1-\theta_e}{d_e}+\frac{1}{2}\log f(r_p(t))\right)
 \cr &&\cr &&
+(t{\rm-independent\;divergent\; terms})\,.
\eea
Note that the whole on shell action is independent of $\alpha$ and $\beta$. This is due to the fact 
that we have used the same free parameters for null vectors for both $v>0$ and $v<0$ regions. Of 
course one could have considered different parameters, but it can be shown that the result will not 
change by adding proper counter terms, as that of previous section. 
 
By making use of the differential equation $\frac{dr_p(t)}{dt}=\frac{f(r_p(t))}{2r_p^{z-1}(t)}$ and the 
fact that the divergent terms are time independent, one can compute the time derivative of the 
above expression and arrive at
\bea
\frac{d}{dt}{\cal C}=\frac{1}{\pi }
\frac{d}{dt}I_{\rm WDW}=\frac{2E}{\pi}
\left(1+\frac{1}{2}\,\frac{d_e}{d_e+z-1}\tilde{f}(r_p(t))\log f(r_p(t))\right),
\eea
where 
\be
\tilde{f}(r_p(t))=\frac{r_h^{d_e+z}}
{r_p^{d_e+z}(t)}-1,\;\;\;\;\;\;\;\;\;\; 
E=\frac{V_d}{16\pi G_N}\;\frac{d_e+z-1}{r_h^{d_e+z}}\,,
\ee
with $E$ being the value at which the complexity growth saturates at late times. It is worth noting 
 that for $z=1$ the above result reduces to that in \cite{Moosa:2017yvt} with effective
 dimension $d_e$.  As we 
mentioned before the energy of the final black brane, i.e., $M$, is related to $E$ by \eqref{EM}. 
It is then evident that for $z\neq 1$ the Lloyd's bound given by $2M$ is again violated and 
the rate of 
complexity growth saturates at $r_p=r_h$ to $2E$. Of course unlike the two sided black brane 
considered in the previous section the ``would be" bound $2E$ is respected in this case. Note 
that 
for $z=1$ when the system is isotropic the growth saturates the Lloyd's bound. Also note that for 
$z\neq 1$ the growth reaches the Lloyd's bound $2M$ at a finite time which is of order of $r_h$.

\begin{figure}
\begin{center}
\includegraphics[scale=0.82]{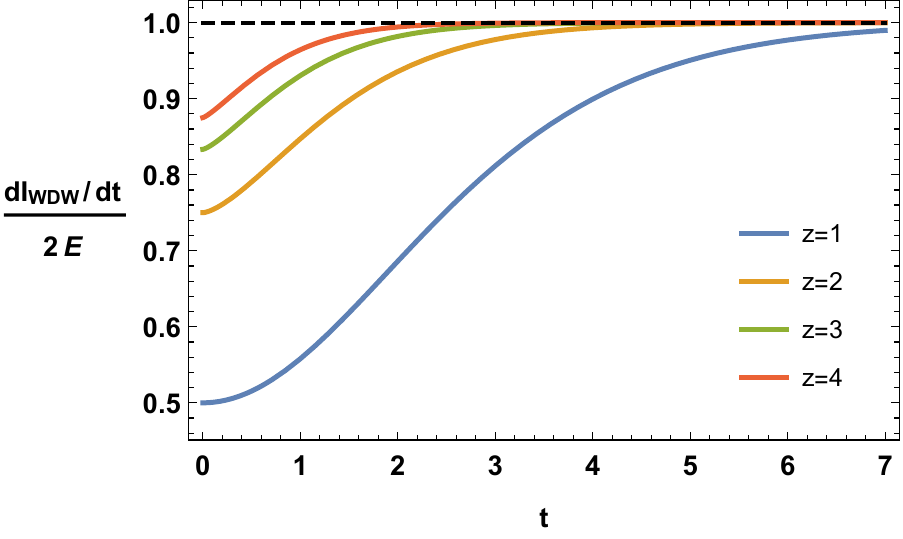}
\hspace{5mm}
\includegraphics[scale=0.82]{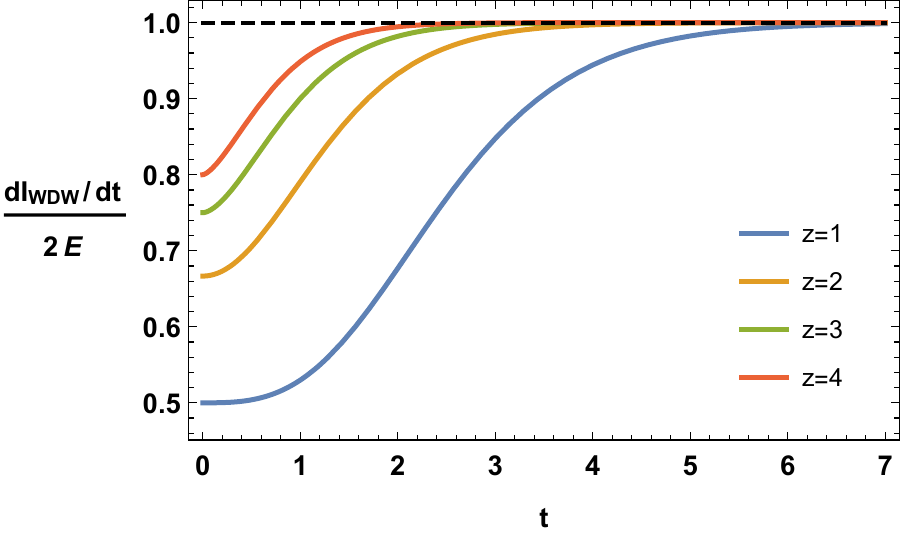}
\end{center}
\caption{Rate of the complexity growth in a WDW patch for Vaidya geometry with $d_e=1$ (left) 
and $d_e=2$ (right).}
\label{fig:D}
\end{figure}

On the other hand at early time where $r_p\sim \delta$ one finds
\bea
\frac{d}{dt}{\cal C}\sim \frac{2E}{\pi}\left(1-\frac{1}{2}\,\frac{d_e}{d_e+z-1}+
\frac{(z/2)^{1/z}}{4}\,\frac{d_e}{d_e+z-1}\frac{t^{1+\frac{d_e}{z}}}{r_h^{d_e+z}}\right).
\eea
Therefore at early time complexity grows as $t^{1+d_e/z}$. The behavior of rate of the complexity 
growth is depicted in Fig.\ref{fig:D} which is consistent with the above time dependence  growth. 
In particular for large $z$ complexity grows linearly with time in the early time regime.

\section{Conclusions}

In this paper we have studied complexity growth, using ``complexity=action'' proposal for a 
gravitational theory admitting anisotropic black brane solution with non-trivial Lifshitz and 
hyperscaling violating exponents. We have considered both one and two sided black brane 
solutions. The two sided black brane would provide a gravitational description for an anisotropic 
thermofiled double state with hyperscaling violation, while the one sided black brane provides a 
gravitational description for a global quench in a hyperscaling violating theory.

We have seen that in both cases the Lloyd's bound is violated given the fact that at late times the 
rate of complexity growth approaches a constant value which is greater than twice the mass of 
the corresponding black brane. Based on this observation we defined a ``would be" bound which 
is the same as Lloyd's bound but the bound is replaced by the new saturation value. It is however 
important to note that in the late time regime in one sided black branes this ``would be" bound is 
approached from below, though for that of two sided black branes it is approached from above. 
This, in turns, means that two sided black brane violates even this ``would be" bound. It is also 
worth noting that for the isotropic case ($z=1$) the ``would be" bound reduces to  Lloyd's bound 
given by twice of the mass.

As far as the divergent terms are concerned the complexity we have obtained has the following
form
\be
\pi {\cal C}=I=-\frac{V_d}{2\pi G_N}\left(\frac{\log\alpha\beta 
\delta^{2(z-\theta_e)}}{2\delta^{d_e}}
+\frac{1-\theta_e}{ d_e \delta^{d_e}}\right)+ {\rm finite\, terms},
\ee
which for $\theta=0, z=1$ reduces to the complicity associated with an $AdS_{d+2}$ 
geometry \cite{Reynolds:2016rvl}.  It is worth noting that from the above volume
behavior of the  complexity, it looks as the theory lives in an effective dimension $d_e$.
This is of course consistent with the holographic  entanglement entropy of 
the theory where we have  $S\sim \frac{{\rm Area}}{\delta^{d_e-1}}$ (see for example 
\cite{Alishahiha:2012cm}).

In order to remove the ambiguity associated with the normalization of null vectors one needs to 
add certain counter terms whose divergent terms are given by
\be
I^{\rm ct}=\frac{V_d}{2\pi G_N}\left(\frac{\log\alpha\beta 
\delta^{2(z-2\theta_e)}}{2\delta^{d_e}}
+\frac{z-2\theta_e}{ d_e \delta^{d_e}}\right)\,.
\ee
Of course these counter terms are not enough to make the complexity finite. We have seen 
that there are other counter terms whose divergent terms are
\be
\tilde{I}^{\rm ct}=\frac{V_d}{2\pi G_N}\left(\frac{\log
\delta^{2\theta_e}}{2\delta^{d_e}}
+\frac{\theta_e-z+1}{ d_e \delta^{d_e}}\right)\,.
\ee
It is then clear that all divergences will be dropped resulting to a finite complexity. It is,
however, important to note that  adding the corresponding counter 
terms  could also have contributions to the finite part of complexity. In 
particular we have seen that this contribution smooths the early time behavior for large $z$ in two 
sided black branes.   This in turns indicates 
the importance of the counter terms
needed to make the action finite. It would be interesting to explore this point more precisely 
\cite{TBA}.

We have also seen that the rate of complexity growth reaches the bound at a finite time that is of 
order of $r_h$.  We have also  evaluated the spatial point $r_m$ associated with this time.
In particular setting $\theta=0, z=1$, one gets 
\be
r_m=2^{\frac{1}{d+1}} r_h.
\ee
Interestingly enough it is the maximal surface that an extremal co-dimension one hypersurface 
inside the black brane could reach at late times \cite{Carmi:2017jqz}.

This means that complexity growth could reach the Lloyd's bound at some time scale comparable 
to thermalization time of a system, but keep evolving non-trivially after that and finally saturate to 
the bound in late times. It would be very interesting to explore this behavior more precisely both 
from gravity and field theory points of view. For recent progress of field theory aspects of 
complexity see \cite{{Jefferson:2017sdb},{Chapman:2017rqy},{Kim:2017qrq},{Khan:2018rzm}}.

\subsection*{Acknowledgements}
The authors would like to kindly thank M.H. Vahidinia and F. Omidi for useful comments  and 
discussions and  also B. Swingle for correspondence. We would also like to thank A. Naseh,  A. 
Shirzad and M. R. Tanhayi for discussions on related topics. We would also like to thank 
referee for his/her comments.
%%%%%%%%%%%%%%%%%%%%%%%%%%%%%%%%%%%%%%%%%
%%%%%%%%%%%%%%%%%%%%%%%%%%%%%%%%%%%%%%%%


\begin{thebibliography}{} 


 %\cite{Ryu:2006bv}
\bibitem{Ryu:2006bv}
 S.~Ryu and T.~Takayanagi,
  ``Holographic derivation of entanglement entropy from AdS/CFT,''
  Phys.\ Rev.\ Lett.\  {\bf 96}, 181602 (2006)
  [hep-th/0603001].
 %CITATION = HEP-TH/0603001;%%


%\cite{Susskind:2014rva}{Stanford:2014jda}\cite{Brown:2015bva}\cite{Brown:2015lvg}
\bibitem{Susskind:2014rva}
  L.~Susskind,
  ``Computational Complexity and Black Hole Horizons,''
  Fortsch.\ Phys.\  {\bf 64}, 24 (2016)
  doi:10.1002/prop.201500092
  [arXiv:1403.5695 [hep-th], arXiv:1402.5674 [hep-th]].
  %%CITATION = doi:10.1002/prop.201500092;%%
  %64 citations counted in INSPIRE as of 17 Dec 2016

%\cite{Stanford:2014jda}
\bibitem{Stanford:2014jda}
  D.~Stanford and L.~Susskind,
  ``Complexity and Shock Wave Geometries,''
  Phys.\ Rev.\ D {\bf 90}, no. 12, 126007 (2014)
  doi:10.1103/PhysRevD.90.126007
  [arXiv:1406.2678 [hep-th]].
  %%CITATION = doi:10.1103/PhysRevD.90.126007;%%
  %47 citations counted in INSPIRE as of 17 Dec 2016


%\cite{Brown:2015bva}\cite{Brown:2015lvg}
\bibitem{Brown:2015bva}
  A.~R.~Brown, D.~A.~Roberts, L.~Susskind, B.~Swingle and Y.~Zhao,
  ``Holographic Complexity Equals Bulk Action?,''
  Phys.\ Rev.\ Lett.\  {\bf 116}, no. 19, 191301 (2016)
  doi:10.1103/PhysRevLett.116.191301
  [arXiv:1509.07876 [hep-th]].
  %%CITATION = doi:10.1103/PhysRevLett.116.191301;%%
  %43 citations counted in INSPIRE as of 17 Dec 2016

 %\cite{Brown:2015lvg}
\bibitem{Brown:2015lvg}
  A.~R.~Brown, D.~A.~Roberts, L.~Susskind, B.~Swingle and Y.~Zhao,
  ``Complexity, action, and black holes,''
  Phys.\ Rev.\ D {\bf 93}, no. 8, 086006 (2016)
  doi:10.1103/PhysRevD.93.086006
  [arXiv:1512.04993 [hep-th]].
  %%CITATION = doi:10.1103/PhysRevD.93.086006;%%
  %28 citations counted in INSPIRE as of 17 Dec 2016



%\cite{Alishahiha:2015rta}{Ben-Ami:2016qex}{Couch:2016exn}{Carmi:2016wjl}
 \bibitem{Alishahiha:2015rta}
  M.~Alishahiha,
  ``Holographic Complexity,''
  Phys.\ Rev.\ D {\bf 92}, no. 12, 126009 (2015)
  doi:10.1103/PhysRevD.92.126009
  [arXiv:1509.06614 [hep-th]].
  %%CITATION = doi:10.1103/PhysRevD.92.126009;%%
  %9 citations counted in INSPIRE as of 03 May 2016

%\cite{Ben-Ami:2016qex}{Couch:2016exn}{Carmi:2016wjl}
\bibitem{Ben-Ami:2016qex} 
  O.~Ben-Ami and D.~Carmi,
  ``On Volumes of Subregions in Holography and Complexity,''
  JHEP {\bf 1611}, 129 (2016)
  doi:10.1007/JHEP11(2016)129
  [arXiv:1609.02514 [hep-th]].
  %%CITATION = doi:10.1007/JHEP11(2016)129;%%
  %9 citations counted in INSPIRE as of 15 Feb 2017



%\cite{Couch:2016exn}{Carmi:2016wjl}
\bibitem{Couch:2016exn}
  J.~Couch, W.~Fischler and P.~H.~Nguyen,
  ``Noether charge, black hole volume and complexity,''
  arXiv:1610.02038 [hep-th].
  %%CITATION = ARXIV:1610.02038;%%
  %5 citations counted in INSPIRE as of 17 Dec 2016


%\cite{Carmi:2016wjl}
\bibitem{Carmi:2016wjl}
  D.~Carmi, R.~C.~Myers and P.~Rath,
  ``Comments on Holographic Complexity,''
  arXiv:1612.00433 [hep-th].
  %%CITATION = ARXIV:1612.00433;%%
  %2 citations counted in INSPIRE as of 17 Dec 2016


%\cite{Bakhshaei:2017qud}
\bibitem{Bakhshaei:2017qud} 
  E.~Bakhshaei, A.~Mollabashi and A.~Shirzad,
``Holographic Subregion Complexity for Singular Surfaces,''
  Eur.\ Phys.\ J.\ C {\bf 77}, no. 10, 665 (2017)
  doi:10.1140/epjc/s10052-017-5247-1
  [arXiv:1703.03469 [hep-th]].


\bibitem{Lloyd:2000}
S.~ Lloyd, ``Ultimate physical limits to computation,''  Nature {\bf 406} (2000) 1047,
[arXiv:quant-ph/9908043]

%\cite{Carmi:2017jqz}
\bibitem{Carmi:2017jqz} 
  D.~Carmi, S.~Chapman, H.~Marrochio, R.~C.~Myers and S.~Sugishita,
  ``On the Time Dependence of Holographic Complexity,''
  JHEP {\bf 1711}, 188 (2017)
  doi:10.1007/JHEP11(2017)188
  [arXiv:1709.10184 [hep-th]].
  %%CITATION = doi:10.1007/JHEP11(2017)188;%%
  %18 citations counted in INSPIRE as of 02 Feb 2018


%\cite{Ghodrati:2017roz,Kim:2017qrq,Moosa:2017yiz}
\bibitem{Ghodrati:2017roz} 
  M.~Ghodrati,
``Complexity growth in massive gravity theories, the effects of chirality, and more,''
  Phys.\ Rev.\ D {\bf 96}, no. 10, 106020 (2017)
  doi:10.1103/PhysRevD.96.106020
  [arXiv:1708.07981 [hep-th]].
  %%CITATION = doi:10.1103/PhysRevD.96.106020;%%
  %5 citations counted in INSPIRE as of 25 Feb 2018

%\cite{Kim:2017qrq}
\bibitem{Kim:2017qrq} 
  R.~Q.~Yang, C.~Niu, C.~Y.~Zhang and K.~Y.~Kim,
``Comparison of holographic and field theoretic complexities for time dependent thermofield double states,''
  JHEP {\bf 1802}, 082 (2018)
  doi:10.1007/JHEP02(2018)082
  [arXiv:1710.00600 [hep-th]].
  %%CITATION = doi:10.1007/JHEP02(2018)082;%%
  %5 citations counted in INSPIRE as of 25 Feb 2018

%\cite{Moosa:2017yiz}
\bibitem{Moosa:2017yiz} 
  M.~Moosa,
``Divergences in the rate of complexification,''
  arXiv:1712.07137 [hep-th].
  %%CITATION = ARXIV:1712.07137;%%
  %2 citations counted in INSPIRE as of 25 Feb 2018

%\cite{Moosa:2017yvt}
\bibitem{Moosa:2017yvt} 
  M.~Moosa,
``Evolution of Complexity Following a Global Quench,''
  arXiv:1711.02668 [hep-th].
  %%CITATION = ARXIV:1711.02668;%%
  %3 citations counted in INSPIRE as of 15 Jan 2018

%\cite{Gouteraux:2011ce}\cite{Huijse:2011ef}
\bibitem{Gouteraux:2011ce}
  B.~Gouteraux and E.~Kiritsis,
  ``Generalized Holographic Quantum Criticality at Finite Density,''
  JHEP {\bf 1112} (2011) 036
  [arXiv:1107.2116 [hep-th]].
  %%CITATION = ARXIV:1107.2116;%%


%\cite{Huijse:2011ef}
\bibitem{Huijse:2011ef} 
  L.~Huijse, S.~Sachdev and B.~Swingle,
  ``Hidden Fermi surfaces in compressible states of gauge-gravity duality,''
  arXiv:1112.0573 [cond-mat.str-el].
  %%CITATION = ARXIV:1112.0573;%%


%\cite{Alishahiha:2012qu}
\bibitem{Alishahiha:2012qu} 
  M.~Alishahiha, E.~O Colgain and H.~Yavartanoo,
  ``Charged Black Branes with Hyperscaling Violating Factor,''
  JHEP {\bf 1211}, 137 (2012)
  [arXiv:1209.3946 [hep-th]].
  %%CITATION = ARXIV:1209.3946;%%
  %20 citations counted in INSPIRE as of 16 Dec 2013

%\cite{Salvio:2013jia}
\bibitem{Salvio:2013jia} 
  A.~Salvio,
  ``Transitions in Dilaton Holography with Global or Local Symmetries,''
  JHEP {\bf 1303}, 136 (2013)
  doi:10.1007/JHEP03(2013)136
  [arXiv:1302.4898 [hep-th]].
  %%CITATION = doi:10.1007/JHEP03(2013)136;%%
  %25 citations counted in INSPIRE as of 01 Mar 2018





%\cite{Swingle:2017zcd}\cite{An:2018xhv}
\bibitem{Swingle:2017zcd} 
  B.~Swingle and Y.~Wang,
  ``Holographic Complexity of Einstein-Maxwell-Dilaton Gravity,''
  arXiv:1712.09826 [hep-th].
  %%CITATION = ARXIV:1712.09826;%%
  %1 citations counted in INSPIRE as of 27 Jan 2018


%\cite{An:2018xhv}
\bibitem{An:2018xhv} 
  Y.~S.~An and R.~H.~Peng,
  ``The effect of Dilaton on the holographic complexity growth,''
  arXiv:1801.03638 [hep-th].
  %%CITATION = ARXIV:1801.03638;%%



%\cite{Parattu:2015gga}
\bibitem{Parattu:2015gga} 
  K.~Parattu, S.~Chakraborty, B.~R.~Majhi and T.~Padmanabhan,
  ``A Boundary Term for the Gravitational Action with Null Boundaries,''
  Gen.\ Rel.\ Grav.\  {\bf 48}, no. 7, 94 (2016)
  doi:10.1007/s10714-016-2093-7
  [arXiv:1501.01053 [gr-qc]].
  %%CITATION = doi:10.1007/s10714-016-2093-7;%%
  %38 citations counted in INSPIRE as of 13 Apr 2017



 %\cite{Lehner:2016vdi}
\bibitem{Lehner:2016vdi}
  L.~Lehner, R.~C.~Myers, E.~Poisson and R.~D.~Sorkin,
  ``Gravitational action with null boundaries,''
  Phys.\ Rev.\ D {\bf 94} (2016) no.8,  084046
  doi:10.1103/PhysRevD.94.084046
  [arXiv:1609.00207 [hep-th]].
  %%CITATION = doi:10.1103/PhysRevD.94.084046;%%
  %16 citations counted in INSPIRE as of 17 Dec 2016


%\cite{Taylor:2008tg}
\bibitem{Taylor:2008tg} 
  M.~Taylor,
  ``Non-relativistic holography,''
  arXiv:0812.0530 [hep-th].
  %%CITATION = ARXIV:0812.0530;%%
  %418 citations counted in INSPIRE as of 05 May 2018



%\cite{Dong:2012se}
\bibitem{Dong:2012se} 
  X.~Dong, S.~Harrison, S.~Kachru, G.~Torroba and H.~Wang,
  ``Aspects of holography for theories with hyperscaling violation,''
  JHEP {\bf 1206}, 041 (2012)
  doi:10.1007/JHEP06(2012)041
  [arXiv:1201.1905 [hep-th]].
  %%CITATION = doi:10.1007/JHEP06(2012)041;%%
  %233 citations counted in INSPIRE as of 08 Feb 2018

%\cite{Dehghani:2015gza}
\bibitem{Dehghani:2015gza} 
  M.~H.~Dehghani, A.~Sheykhi and S.~E.~Sadati,
  ``Thermodynamics of nonlinear charged Lifshitz black branes with hyperscaling violation,''
  Phys.\ Rev.\ D {\bf 91}, no. 12, 124073 (2015)
  doi:10.1103/PhysRevD.91.124073
  [arXiv:1505.01134 [hep-th]].
  %%CITATION = doi:10.1103/PhysRevD.91.124073;%%
  %20 citations counted in INSPIRE as of 13 Feb 2018


%\cite{Henningson:1998gx}
\bibitem{Henningson:1998gx} 
  M.~Henningson and K.~Skenderis,
  ``The Holographic Weyl anomaly,''
  JHEP {\bf 9807}, 023 (1998)
  doi:10.1088/1126-6708/1998/07/023
  [hep-th/9806087].
  %%CITATION = doi:10.1088/1126-6708/1998/07/023;%%
  %1210 citations counted in INSPIRE as of 12 May 2018





\bibitem{TBA}
A. Akhavan, M. Alsihahiha, F. Omidi, ``Complexity and Counterterms,'' To appear.


%\cite{Alishahiha:2014cwa}
\bibitem{Alishahiha:2014cwa} 
  M.~Alishahiha, A.~F.~Astaneh and M.~R.~M.~Mozaffar,
  ``Thermalization in Backgrounds with Hyperscaling Violating Factor,''
  arXiv:1401.2807 [hep-th].
  %%CITATION = ARXIV:1401.2807;%%
  %1 citations counted in INSPIRE as of 24 Mar 2014


%\cite{Reynolds:2016rvl}
\bibitem{Reynolds:2016rvl} 
  A.~Reynolds and S.~F.~Ross,
  ``Divergences in Holographic Complexity,''
  Class.\ Quant.\ Grav.\  {\bf 34}, no. 10, 105004 (2017)
  doi:10.1088/1361-6382/aa6925
  [arXiv:1612.05439 [hep-th]].
  %%CITATION = doi:10.1088/1361-6382/aa6925;%%
  %30 citations counted in INSPIRE as of 25 Jun 2018
  
%\cite{Alishahiha:2012cm}
\bibitem{Alishahiha:2012cm} 
  M.~Alishahiha and H.~Yavartanoo,
  ``On Holography with Hyperscaling Violation,''
  JHEP {\bf 1211}, 034 (2012)
  doi:10.1007/JHEP11(2012)034
  [arXiv:1208.6197 [hep-th]].
  %%CITATION = doi:10.1007/JHEP11(2012)034;%%
  %49 citations counted in INSPIRE as of 25 Jun 2018
  
  
  


%\cite{Jefferson:2017sdb}{Chapman:2017rqy}{Kim:2017qrq}{Khan:2018rzm}
\bibitem{Jefferson:2017sdb} 
  R.~Jefferson and R.~C.~Myers,
  ``Circuit complexity in quantum field theory,''
  JHEP {\bf 1710}, 107 (2017)
  doi:10.1007/JHEP10(2017)107
  [arXiv:1707.08570 [hep-th]].
  %%CITATION = doi:10.1007/JHEP10(2017)107;%%
  %17 citations counted in INSPIRE as of 01 Mar 2018



%\cite{Chapman:2017rqy}{Kim:2017qrq}{Khan:2018rzm}
\bibitem{Chapman:2017rqy} 
  S.~Chapman, M.~P.~Heller, H.~Marrochio and F.~Pastawski,
  ``Towards Complexity for Quantum Field Theory States,''
  arXiv:1707.08582 [hep-th].
  %%CITATION = ARXIV:1707.08582;%%
  %16 citations counted in INSPIRE as of 01 Mar 2018


%\cite{Khan:2018rzm}
\bibitem{Khan:2018rzm} 
  R.~Khan, C.~Krishnan and S.~Sharma,
  ``Circuit Complexity in Fermionic Field Theory,''
  arXiv:1801.07620 [hep-th].
  %%CITATION = ARXIV:1801.07620;%%



\end{thebibliography}
\end{document}